# Dual-Material Double-Gate Source-Pocket Tunnel Field Effect Transistor with Homogeneous Gate Dielectric: Computational Analysis of Structural and Material Parameters for Enhanced Performance

Ramisa Fariha, Saikat Das, Labiba Tanjil Nida, Abeer Khan, and Md Tashfiq Bin Kashem*

Department of Electrical and Electronic Engineering, Ahsanullah University of Science and Technology, Dhaka-1208, Bangladesh.

*Corresponding author. E-mail: tashfiq.eee@aust.edu

Contributing authors: ramisafariha11@gmail.com; dsaikat952@gmail.com; labibatanjilnida12@gmail.com; abeerkhan.2100@gmail.com

All authors contributed equally to this work.

**Abstract**

Dual-material double-gate tunnel field effect transistor (DMDG TFET) is a promising candidate for low-power, high-speed electronics due to enhanced electrostatic control and superior switching characteristics. Integrating a pocket region between the source and channel—doped oppositely to the source—further improves tunneling efficiency by modulating the electric field at the tunneling junction. This combined architecture, termed the DMDG source-pocket TFET (DMDG-SP TFET), achieves higher ON current and reduced subthreshold swing compared to conventional TFETs. Previous DMDG-SP TFET designs primarily use heterogeneous gate dielectrics, composed of two stacked insulators to enhance gate control and tunneling modulation. However, such hetero gate dielectrics increase fabrication complexity and may degrade device reliability due to material incompatibility. This work proposes a silicon-based DMDG-SP TFET employing a homogenous gate dielectric, investigated through Silvaco Atlas-based 2-D TCAD simulations, aiming to simplify fabrication without compromising performance. Presence of the pocket results in 6.7× higher ON current and 1.7× lower subthreshold swing compared to pocket-less devices. Dual-material gates boost ON current by 45% and improve the ON/OFF current ratio by 59% compared to single-material gates in pocket-based devices. Detailed simulations analyze effects of gate metal work functions and lengths, gate dielectric constant, and doping densities and lengths of all regions. The optimized device achieves an ON current of $3.16\times10^{-4}$ A/μm, OFF current of $1.54\times10^{-17}$ A/μm, ON/OFF ratio of $2.05\times10^{13}$, and subthreshold slope of 6.29 mV/decade. These findings offer critical insights for designing manufacturable, high-performance homojunction silicon-based DMDG-SP TFETs with homogeneous gate dielectrics for next-generation low-power integrated circuits.

**Keywords: Tunnel FET, Dual-material, Double-gate, Source-pocket, Homogeneous gate dielectric**

# 1 Introduction

The progressive development of metal-oxide-semiconductor field effect transistor (MOSFET) has resulted in remarkable enhancements in integrated circuit technology, characterized by miniaturization, as well as increased speed and energy efficiency over the past several decades [1, 2]. However, MOSFET faces some inherent limitations: it consumes considerable power, particularly in standby mode, and is constrained by a fundamental subthreshold swing limit of 60 mV/decade at room temperature. This restriction limits ultra-low power operation, while the increasing short-channel effects at smaller technology nodes exacerbate leakage currents, undermining device performance [3].

Tunnel FET (TFET) has emerged as promising alternative to conventional MOSFET for low-power applications since its introduction in 1992 [4–6]. TFET achieves substantially lower subthreshold swing (< 60 mV/decade) through band-to-band tunneling mechanisms, enabling superior control of ON/OFF states and reducing power consumption [7–9]. Unlike MOSFET that depends on thermionic emission, TFET allows carriers to tunnel through a potential barrier under a small applied voltage, mitigating short-channel effects and allowing faster switching at lower voltages [4, 10].

Despite these advantages, TFET suffers from inherently low ON current due to quantum tunneling-based carrier injection, limiting its drive capabilities and restricting applicability in high-performance circuits. The tunneling efficiency in TFET depends on both material properties and device architecture. Poor tunneling control can lead to insufficient ON current and higher OFF current, making it difficult to achieve optimal performance [11].

To address these challenges, the double-gate configuration (DG TFET) has been proposed, employing two gates to exert enhanced electrostatic control over the channel. DG TFET exhibits significantly increased ON current while maintaining ultra-low OFF current levels (in the femtoampere to picoampere range), improving short-channel behavior and ON/OFF current ratios [12–14].

Building on this, the introduction of dual-material gates in DG TFET, abbreviated as DMDG TFET, further refines device performance [15]. By integrating two metals with different work functions in the gate structure, the electric field, and surface potential along the channel can be engineered to simultaneously optimize ON/OFF currents, threshold voltage, and subthreshold slope [15–19].

Another effective design enhancement involves the use of a source-pocket region—an area between the source and channel—doped oppositely to the source. This pocket modifies the electric field distribution, increasing carrier injection at the tunneling junction, thereby boosting ON current and steepening the subthreshold slope while reducing OFF-state leakage [11, 20–27].

Previous studies on DG TFET with source-pocket primarily focused on either single-material gate structures [11, 20, 28] or dual-material gates combined with hetero-dielectric gate stacks [29–31]. Although hetero-dielectric gate insulators—combining high-*k* and low-*k* materials—offer enhanced control over tunneling, their complex fabrication and material compatibility issues pose significant challenges for large-scale manufacturing. Issues such as thermal expansion mismatch and interface traps can degrade device performance by increasing leakage currents and reducing reliability [32, 33]. In contrast, using a single homogeneous dielectric as gate insulator simplifies fabrication while still allowing the benefits of multi-material gate structures and pocket engineering to be harnessed [34].

So far, only a limited number of studies have examined DMDG TFETs with single homogeneous gate dielectric; however, these did not incorporate pocket regions [17–19]. To the best of the authors' knowledge, no study has yet explored the combined benefits of DMDG features and source-pocket engineering in a single-gate-dielectric-based structure.

In this paper, a silicon (Si)-based DMDG TFET with a source-pocket (DMDG-SP TFET) employing a single homogeneous gate dielectric has been investigated. A comprehensive computational analysis of its device performance has been conducted, along with comparisons to existing TFET configurations.

The remainder of this paper is organized as follows. Section 2 details the device structure, and key design parameters. Section 3 describes the simulation models and methodology used. Section 4 presents simulation results, including performance comparisons, and parametric analyses. Finally, Section 5 concludes with a summary of key findings.

# 2 Device structure

Figure 1 shows the design of the homojunction Si-based DMDG-SP TFET. The device has a p–n–p–n structure: the source is heavily p-doped, the intermediate pocket between source and channel is heavily n-doped, the channel is lightly p-doped, and the drain is moderately n-doped. Uniform doping profile is assumed across all regions. The gate consists of two metal electrodes with different work functions: the metal near the source, called the tunneling gate, controls the tunneling of electrons at the source-channel junction, while the metal near the drain, called the auxiliary gate, can suppress the ambipolar transport. The gate insulator is made of hafnium oxide ($HfO_2$), a high-*k* dielectric material. All the dimensions and material parameters used in the simulations are listed in

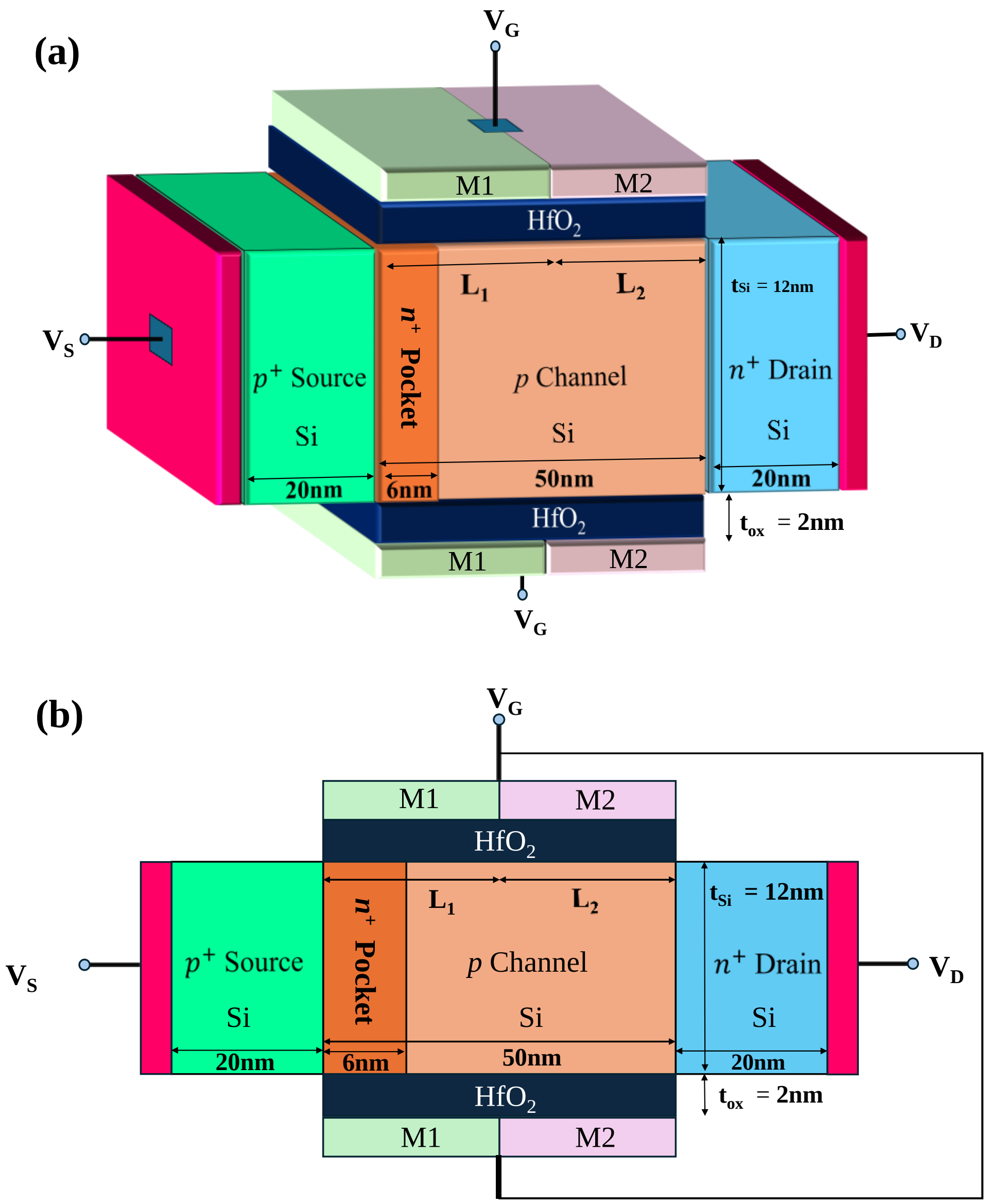

**Fig. 1** Schematic of the Si-based homojunction DMDG-SP TFET with homogeneous gate dielectric: (a) 3-D model and (b) 2-D version as utilized in the simulations of this work.

Table 1, unless explicitly stated otherwise. The 1-D energy band diagrams presented in the Results and Discussion section correspond to horizontal cross-sections along the channel, positioned 1 nm below the gate insulator–channel interface.

# 3 Simulation methodology

Silvaco Atlas Technology Computer Aided Design (TCAD) software is utilized for the simulations in this work [35]. Nonlocal band-to-band (BTB) tunneling model, which incorporates the spatial variation of the energy bands, is employed to accurately capture the quantum tunneling phenomena at the source-channel and drain-channel interfaces. To account for bandgap narrowing effects at high doping densities, the bandgap narrowing (BGN) model is integrated. Besides, mobility models dependent on electric field and impurity concentration, along with Shockley-Read-Hall (SRH) and Auger recombination mechanisms, are included to ensure precise prediction of device behavior.

**Table 1** Device parameters used in the simulations

| Symbol | Parameter | Value |
|---|---|---|
| $L_g$ | Gate length | 50 nm |
| $t_{Si}$ | Silicon film thickness | 12 nm |
| $t_{HfO2}$ | $HfO_2$ thickness | 2 nm |
| $N_{channel}$ | Channel doping | $10^{16}$ $cm^{-3}$ |
| $N_{source}$ | Source doping | $10^{20}$ $cm^{-3}$ |
| $N_{drain}$ | Drain doping | $5\times10^{18}$ $cm^{-3}$ |
| $\Phi_{M1}$ | Tunneling gate work function | 4.4 eV |
| $\Phi_{M2}$ | Auxiliary gate work function | 4.6 eV |
| $L_1$ | Tunneling gate length | 25 nm |
| $L_2$ | Auxiliary gate length | 25 nm |
| $N_{pocket}$ | Pocket doping | $3\times10^{19}$ $cm^{-3}$ |
| $L_{pocket}$ | Pocket length | 6 nm |
| $L_s$ | Source length | 20 nm |
| $L_d$ | Drain length | 20 nm |

# 4 Results and discussion

## 4.1 Working principle of DMDG-SP TFET

The working mechanism of the DMDG-SP TFET can be explained using the energy band diagrams shown in Fig. 2. In the OFF state, without any applied gate voltage (Fig. 2a), conduction band edge ($E_C$) in the pocket and channel lie above the valence band edge ($E_V$) of the source. The electric field at the source-pocket junction remains low (Fig. 2c), the tunneling barrier remains high, and BTB tunneling of electrons is not possible (Fig. 2e). When a positive gate voltage is applied in the ON state (Fig. 2b), the electric field at the source-pocket junction increases (Fig. 2d), pushing $E_C$ of the pocket downward to go below the $E_V$ of the source. This reduces the tunneling barrier for valence electrons in the source, allowing them to tunnel into the conduction band of the pocket (Fig. 2f). Subsequently the electrons reach the drain by drift-diffusion [28], producing a drain current ($I_D$) from drain to source.

## 4.2 Comparison between DMDG-SP TFET and pocket-free DMDG TFET

The advantages of introducing a heavily n- doped (n+) pocket between the source and the channel in the DMDG TFET can be seen from the analysis results in Fig. 3. The enhanced built-in potential at the p+ source/n+ pocket junction produces a sharp change and local minimum in the $E_C$ (Fig. 3a) [11], resulting in a stronger lateral electric field compared to the pocket-free structure (Fig. 3b). This narrows the tunneling barrier and increases the electron tunneling rate from the source to the pocket (Fig. 3c). Consequently, the pocket-based DMDG TFET exhibits a significantly higher drain current than its pocket-free counterpart, as shown in the transfer and output characteristics in Fig. 4.

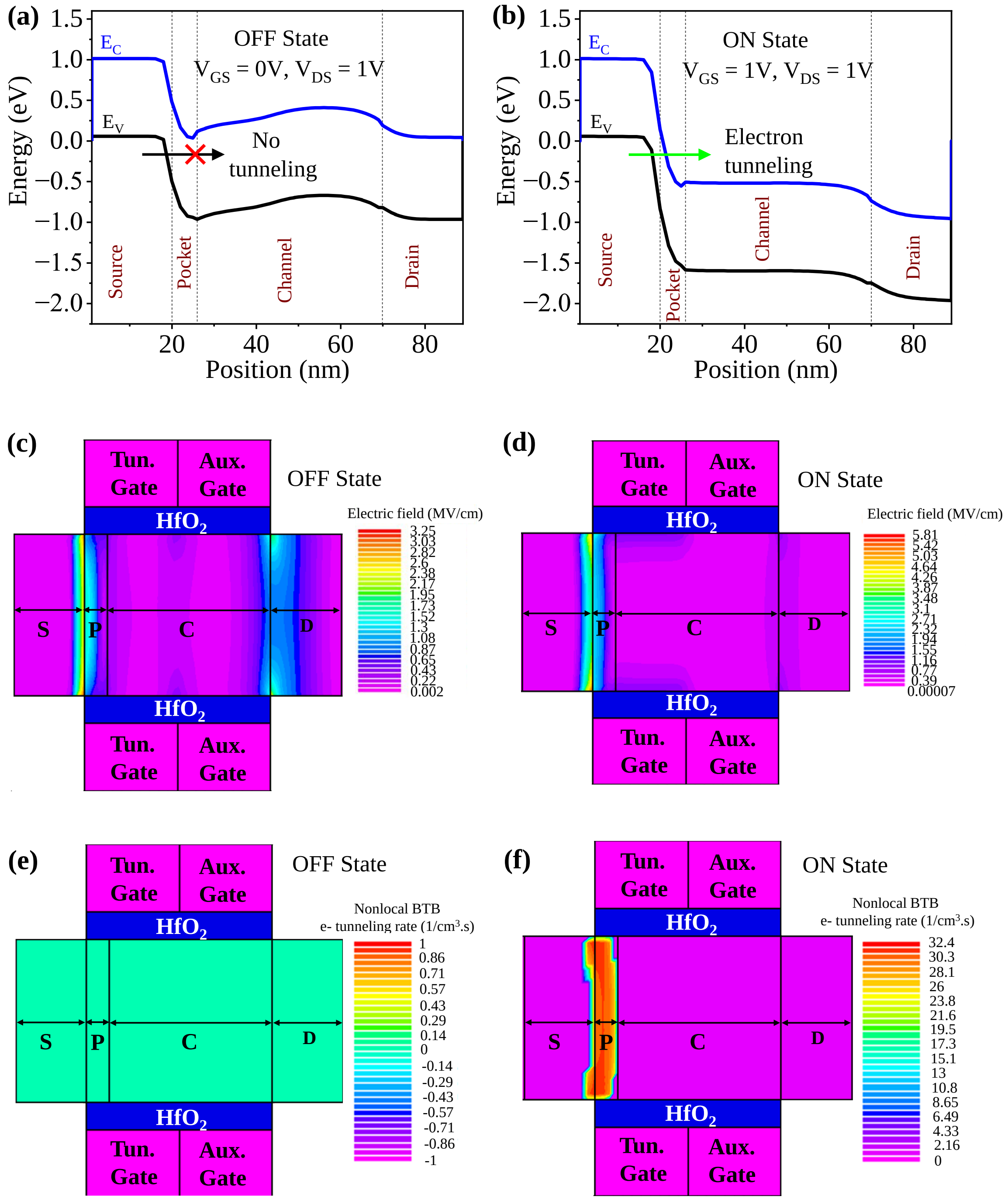

**Fig. 2** Analysis of DMDG-SP TFET in terms of: (a) and (b) energy band diagram in OFF and ON states; (c) and (d) contour of electric field in OFF and ON states; and (e) and (f) contour of electron tunneling rates in OFF and ON states. Tun. and Aux. indicate tunneling and auxiliary respectively. S, P, C and D in the figures stand for source, pocket, channel and drain region respectively.

## 4.3 Comparison of DMDG-SP TFET with other pocket-based TFET structures

Figure 5 compares the DMDG-SP TFET with two other source-pocket based TFETs: single-material single-gate (SMSG-SP) and single-material double-gate (SMDG-SP) devices. The DMDG-SP TFET achieves the highest current over the range of gate and drain bias voltages considered (Fig. 5a, b). At $V_{GS}$ (gate-to-source voltage) = $V_{DS}$ (drain-to-source voltage) = 1V, $I_D$ of DMDG-SP TFET is 1.4 times greater than that of the SMDG-SP TFET and 2.6 times greater than that of the SMSG-SP TFET. This superior performance of the DMDG-SP TFET stems

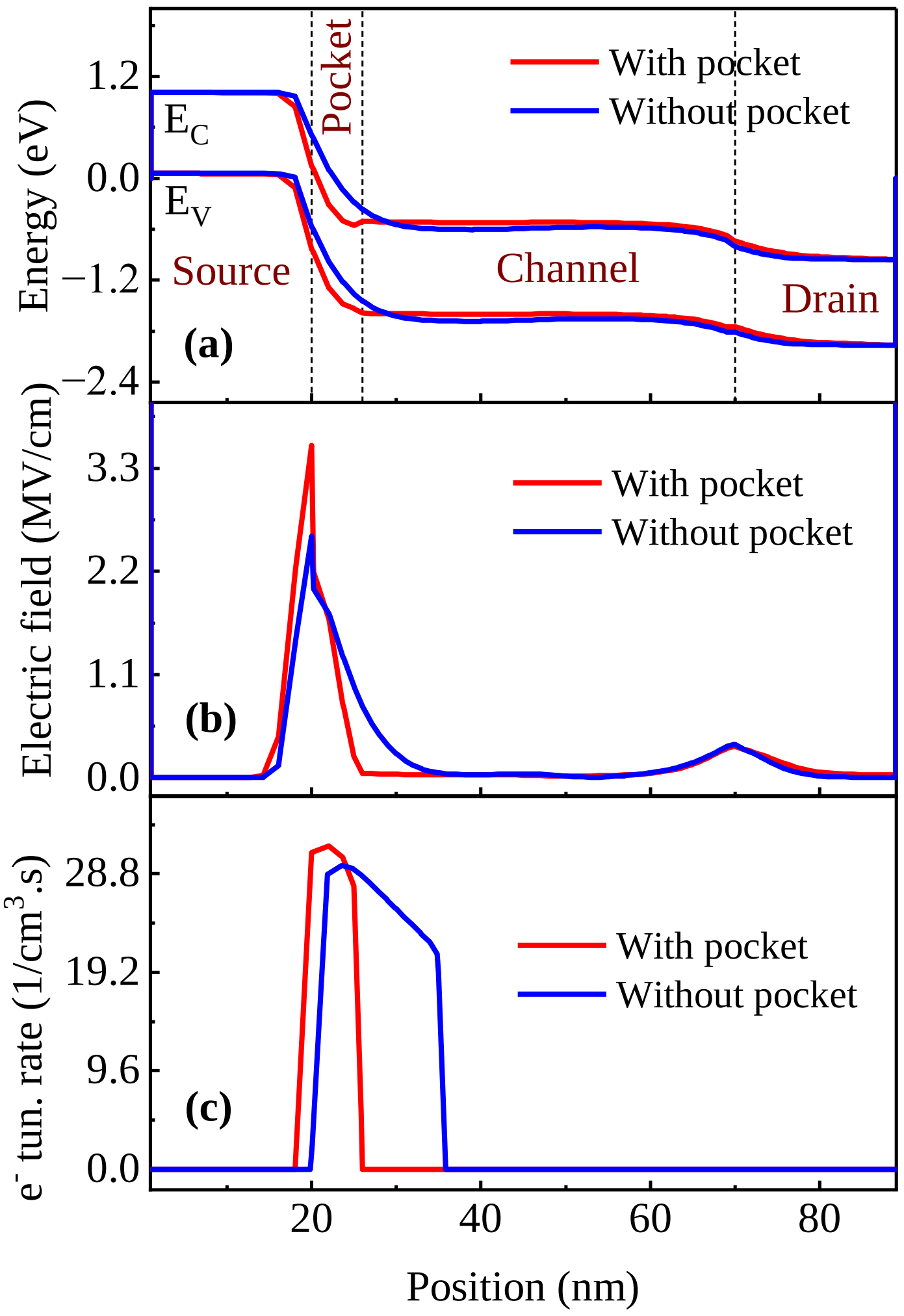

**Fig. 3** Comparison between DMDG TFETs with and without the source-pocket in terms of: (a) energy band diagram, (b) electric field, and (c) nonlocal BTB electron tunneling rate in ON state ($V_{GS} = V_{DS} = 1V$).

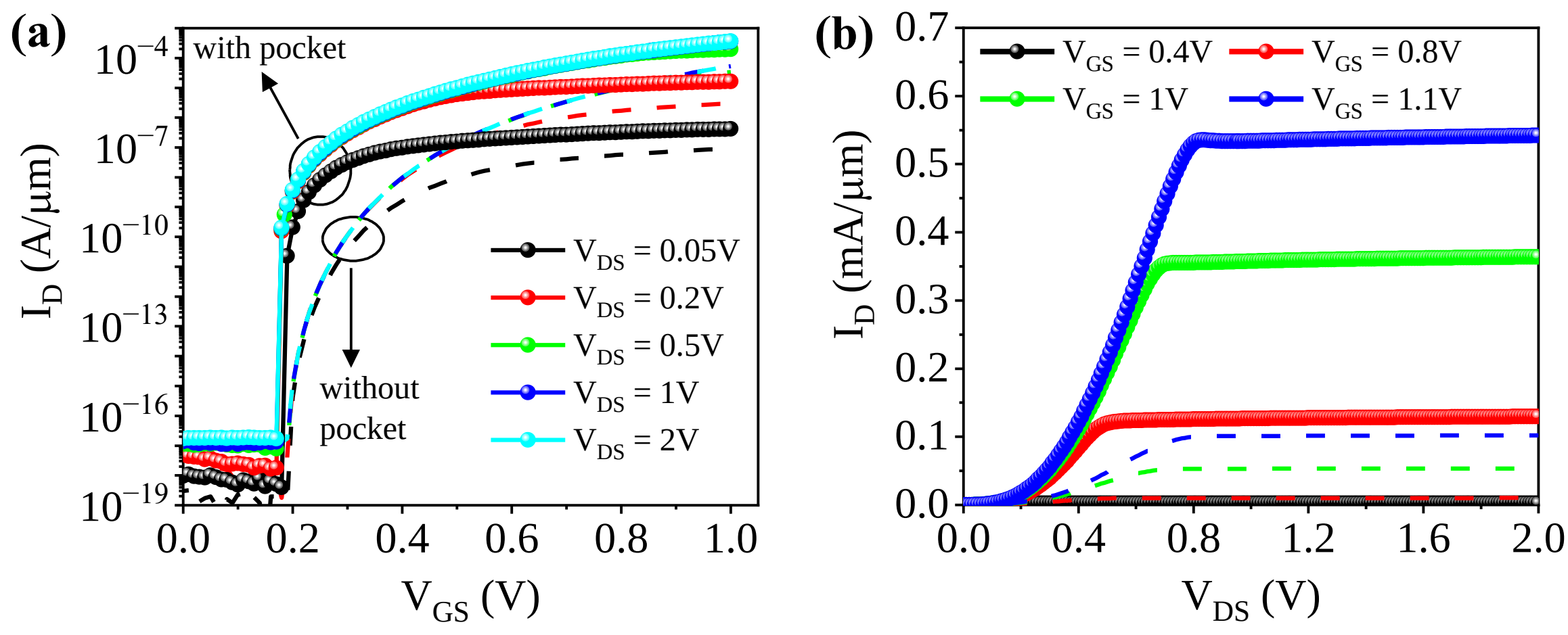

**Fig. 4** (a) Transfer characteristics and (b) output characteristics of DMDG TFET with and without the source-pocket. In both figures, solid lines with symbols represent the device with pocket, and dashed lines represent the device without the pocket region. Higher current can be obtained in pocket-based structure; for example, at $V_{DS}$ = 1V and $V_{GS}$ = 1V, $I_D$ is 6.7 times higher than that of the pocket-free TFET.

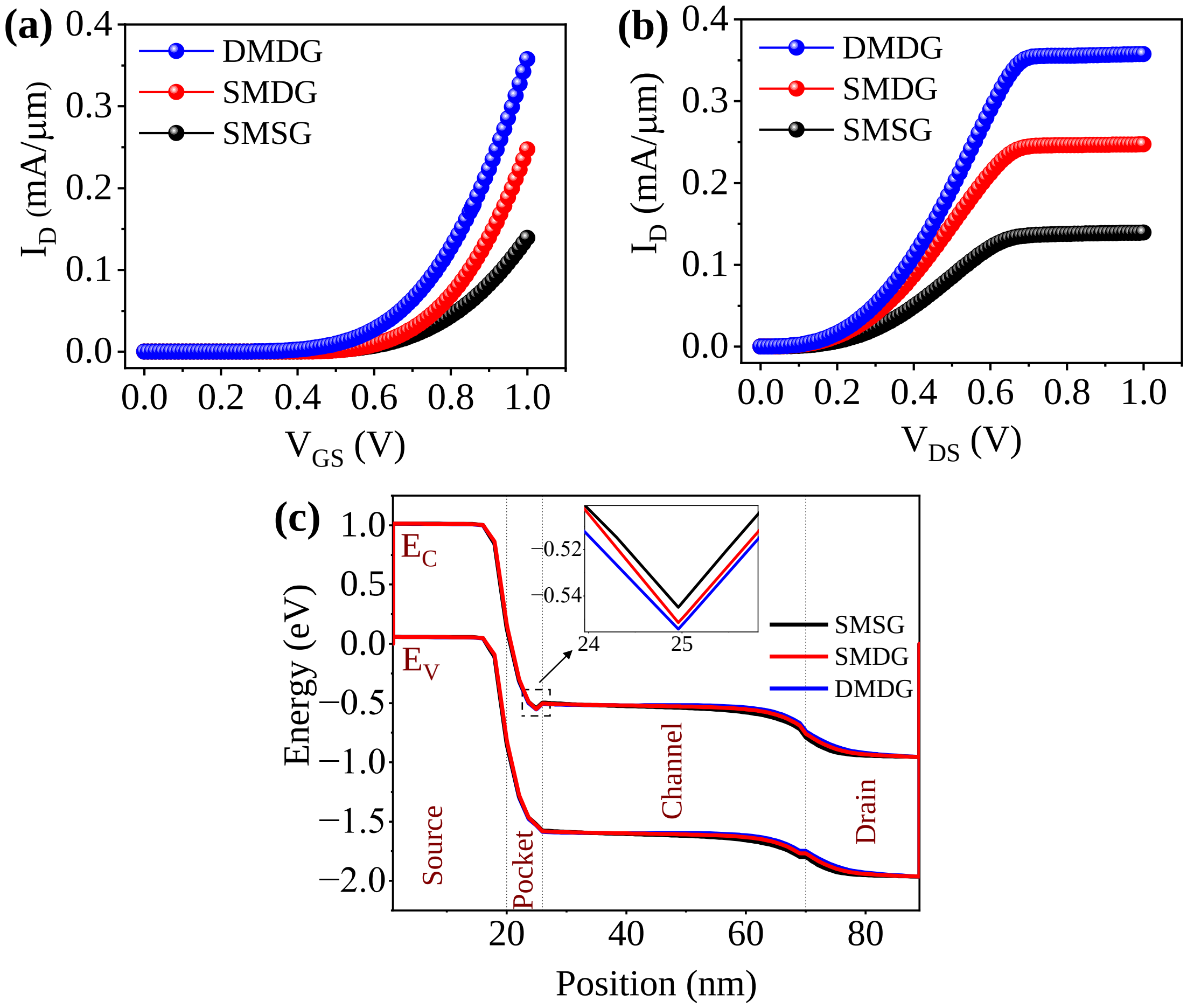

**Fig. 5** Comparison among SMSG, SMDG, and DMDG TFETs with source-pocket in terms of (a) transfer characteristics, (b) output characteristics, and (c) energy band diagram in ON state ($V_{GS}$ = $V_{DS}$ = 1V). All the structures include n+ pocket between source and channel. Metal with work function of 4.6 eV is used as the gate material in single-material gate TFETs.

from larger band bending within the pocket region (Fig. 5c), due to the enhanced gate control, which narrows the tunneling barrier and increases electron tunneling from the source.

## 4.4 Material and structural parameter analysis of DMDG-SP TFET

### 4.4.1 Work function of tunneling and auxiliary gate metals

The effects of the work functions of the tunneling and auxiliary gate metals on the current-voltage characteristics of the DMDG-SP TFET are explored separately (Figs. 6 and 7). First, for the variation only in the tunneling gate metal, threshold voltage increases as the corresponding work-function ($\Phi_{M1}$) increases (Fig. 6a). This occurs because the conduction band shifts upward in the channel under the tunneling gate (Fig. 6b), reducing the probability of tunneling, and consequently increasing the threshold voltage. In contrast, varying the auxiliary gate metal work function ($\Phi_{M2}$) while keeping $\Phi_{M1}$ constant has little effect on the tunnel junction at the source-pocket interface (Fig. 6d) and does not change the threshold voltage (Fig. 6c).

The transfer characteristics show a slight decrease in ON current with increasing $\Phi_{M1}$, whereas an increase in $\Phi_{M2}$ causes a more pronounced reduction (Fig. 7a, c). When $\Phi_{M1}$ varies and $\Phi_{M2}$ is kept constant, $E_C$ in the channel remains almost unchanged in the ON state (Fig. 7b). Therefore, once the device is turned on, the tunneling area for electrons varies little, leading to minimal changes in the ON current (Fig. 7a). Conversely, when $\Phi_{M2}$ is varied, $E_C$ in the channel shifts to higher energy level for higher $\Phi_{M2}$ (Fig. 7d). This reduces the tunneling area for electrons from the valence band of the source and, consequently, decreases the ON current (Fig. 7c).

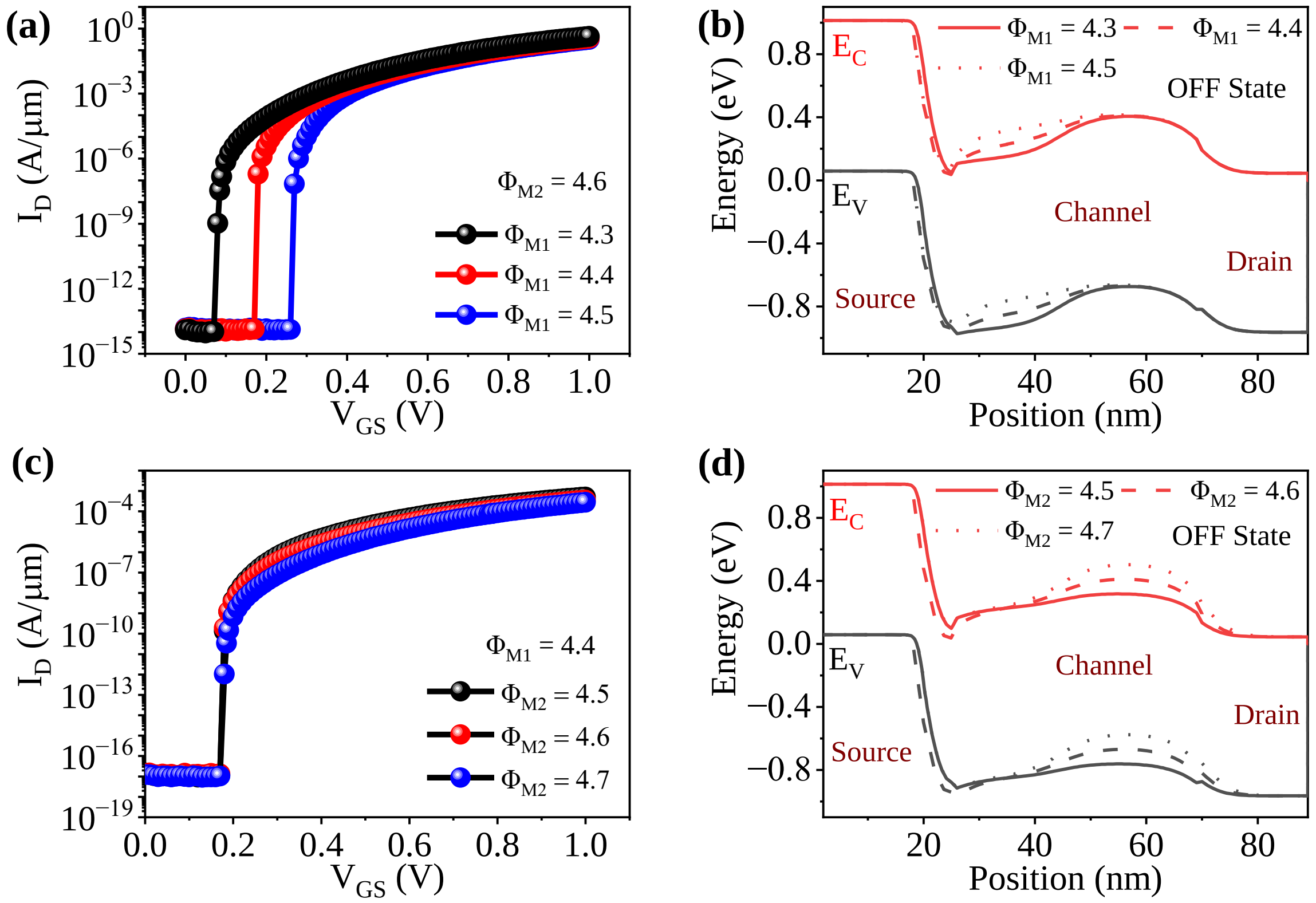

**Fig. 6** (a) Transfer characteristics of DMDG-SP TFET (log scale) for different tunneling gate metal work functions and (b) corresponding energy band diagram in OFF state ($V_{GS}$ = 0V, and $V_{DS}$ = 1V); (c) transfer characteristics of DMDG-SP TFET (log scale) for different auxiliary gate metal work functions and (b) corresponding energy band diagram in OFF state ($V_{GS}$ = 0V, and $V_{DS}$ = 1V). Metals such as Aluminum, Titanium, Chromium, Tungsten, and Silver represent work functions in the 4.3–4.7 eV range.

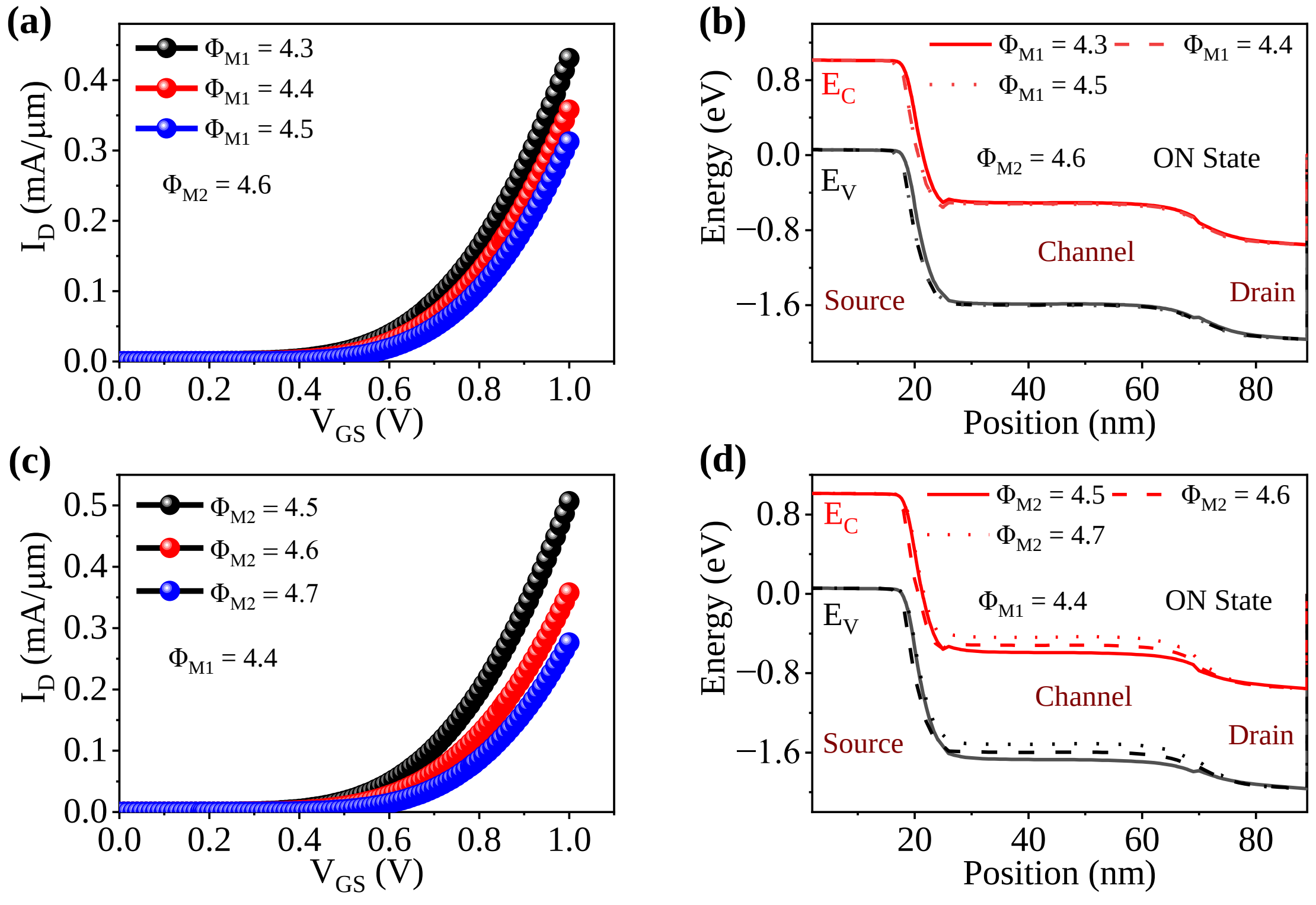

**Fig. 7** (a) Transfer characteristics of DMDG-SP TFET (linear scale) for different tunneling gate metal work functions and (b) corresponding energy band diagram in ON state ($V_{GS}$ = $V_{DS}$ = 1V); (c) transfer characteristics of DMDG-SP TFET (linear scale) for different auxiliary gate metal work functions and (b) corresponding energy band diagram in ON state ($V_{GS}$ = $V_{DS}$ = 1V).

### 4.4.2 Length ratio of tunneling and auxiliary gate metals

This study investigates the impact of the ratio between the tunneling gate length ($L_1$) and auxiliary gate length ($L_2$) for a fixed total gate length of 50 nm (Fig. 8). Specifically, $L_1$ values of 25 nm, 17 nm, and 10 nm correspond to $L_1$:$L_2$ ratios of 1:1, 1:2, and 1:4, respectively. BTB tunneling probability at the source-pocket junction increases with increasing $L_1$ (higher $L_1$:$L_2$ ratio), due to the $E_C$ moving closer to the $E_V$ of the source (Fig. 8a). This shift reduces the tunneling barrier width, thereby lowering the threshold gate voltage required to activate the device, and increasing the ON current [17] (Fig. 8b).

### 4.4.3 Dielectric constant of the gate insulator

The role of the dielectric constant of the gate insulator is investigated in Fig. 9. Gate insulators considered include $SiO_2$, $Si_3N_4$, $Y_2O_3$, and $HfO_2$, with dielectric constants (*k*) of 3.9, 7.5, 15, and 25, respectively. Higher-*k* gate dielectrics enhance the capacitive coupling between the gate electrode and the source-pocket tunnel junction, which intensifies the electric field modulation at the tunneling junction. Consequently, the ON current exhibits a steeper increase with gate voltage due to increased BTB tunneling rates, resulting in higher drain current, and improved ON/OFF current ratio.

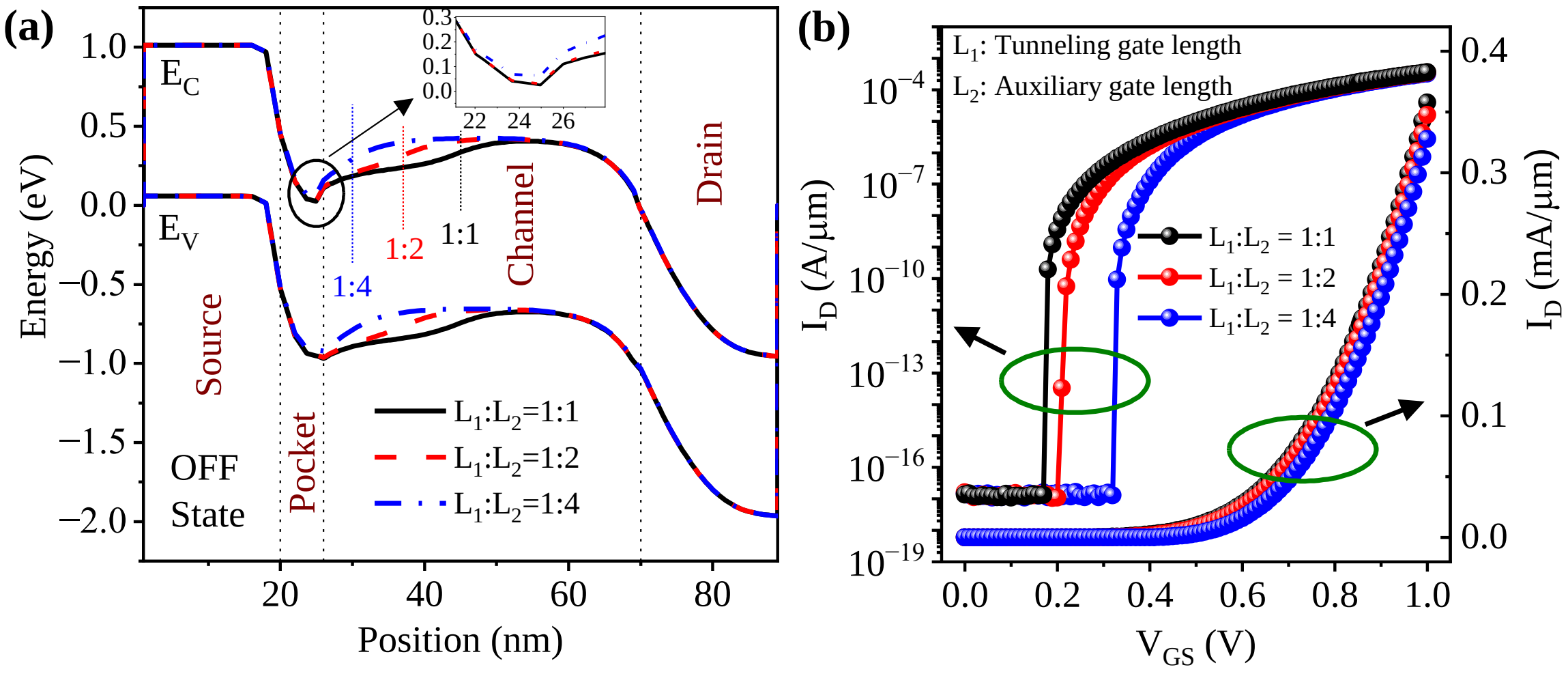

**Fig. 8** Effect of ratio of length of tunneling and auxiliary gate metal on: (a) energy band diagram in OFF state ($V_{GS}$ = 0V, $V_{DS}$ = 1V), and (b) transfer characteristics.

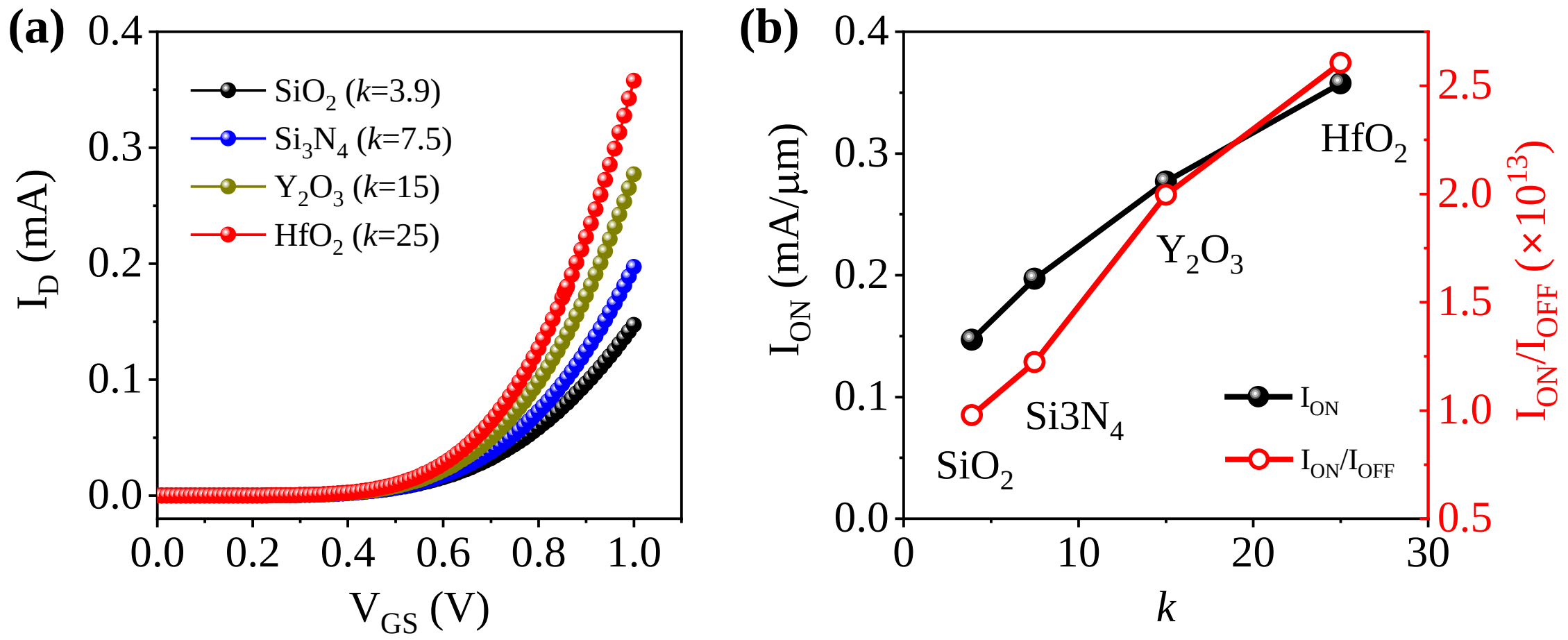

**Fig. 9** Effect of dielectric constant (*k*) of different materials as the gate insulator on: (a) transfer characteristics, and (b) ON current and ON/OFF current ratio.

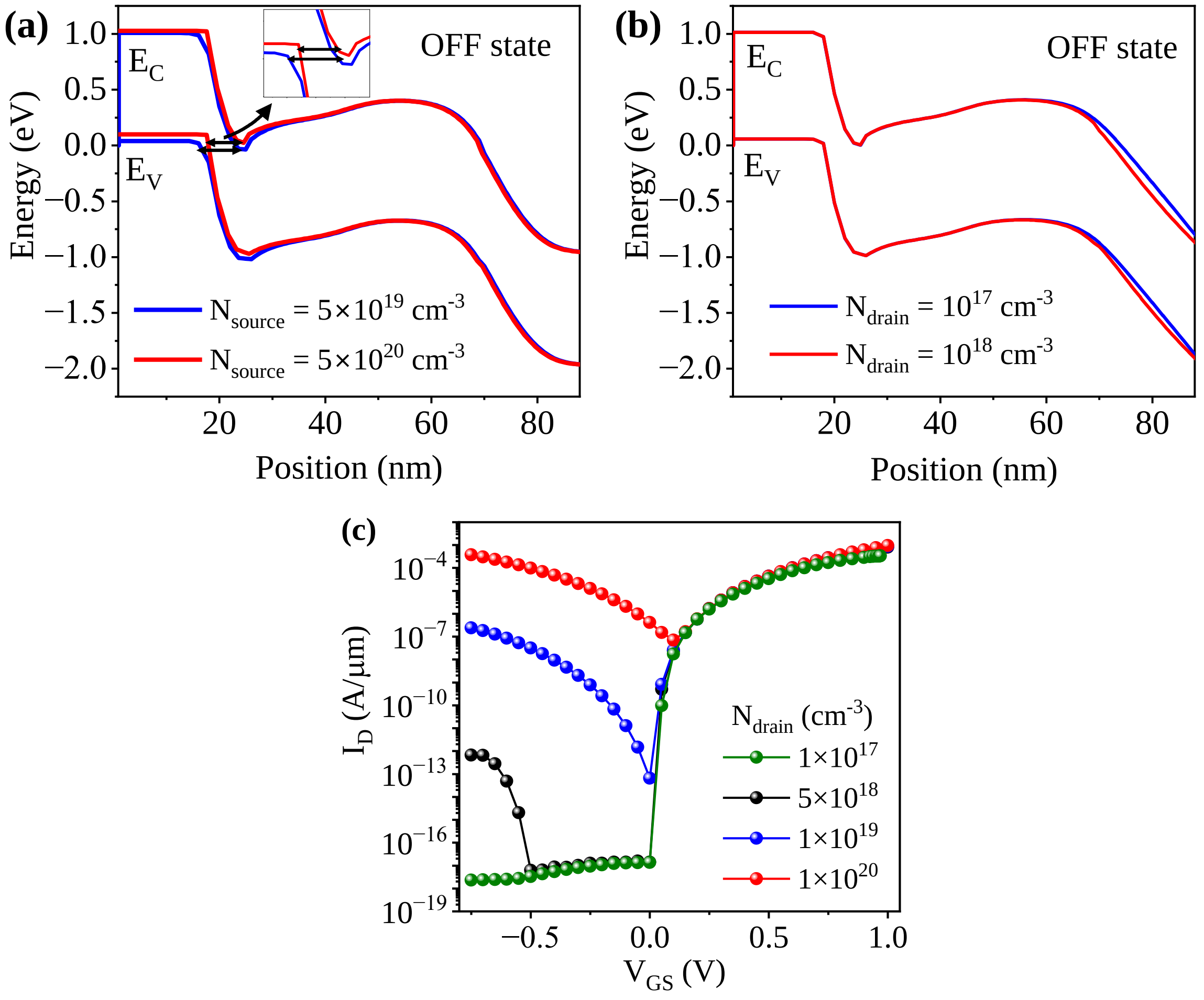

**Fig. 10** (a) and (b) Energy band diagram in OFF state ($V_{GS}$ = 0V, $V_{DS}$ = 1V), and (c) transfer characteristics for different source and drain doping density ($N_{source}$ and $N_{drain}$ respectively). In (a), $N_{drain}$ is kept constant at $5\times10^{18}$ cm$^{-3}$ and in (b) and (c) $N_{source}$ is kept constant at $10^{20}$ cm$^{-3}$. The horizontal arrows in (a) indicate the corresponding tunneling barrier widths.

## 4.4.4 Source and drain doping density

This study investigates the effects of doping density in the source and drain regions by analyzing ON/OFF current and subthreshold slope (SS). Increasing the source doping density reduces the depletion region width and enhances the band bending gradient at the source-pocket junction [36, 37] (Fig. 10a). This results in a narrower tunneling barrier width, thereby increasing BTB tunneling probability, which manifests as higher ON current and reduced subthreshold swing in devices with heavily doped sources (Fig. 11a, d). Although the OFF current exhibits a slight increase with higher source doping due to increased leakage paths, this effect remains minimal (Fig. 11b). Ultimately heavier source doping improves the ON/OFF current ratio (Fig. 11c).

Higher drain doping density steepens the energy band bending at the channel-drain junction (Fig. 10b), which increases ambipolar conduction (Fig. 10c) by facilitating BTB tunneling at the drain side [17]. Analysis of Fig. 11a and b indicates that drain doping density has a minimal influence on the ON current but significantly impacts the OFF current. This behavior is attributed to the highly doped drain causing a majority of the drain bias to drop across the channel region [38], thereby intensifying the band bending at the channel-drain interface and increasing the OFF current. Conversely, a lightly doped drain distributes the drain voltage drop between the drain and channel regions, extending the tunneling barrier into the drain during the OFF state [38]. This results in lower OFF current, leading to enhanced ON/OFF current ratio, and improved subthreshold slope (Fig. 11b–d).

## 4.4.5 Channel doping

The impact of channel doping on device performance is examined for doping concentrations of $10^{16}$, $10^{17}$ and $10^{18}$ cm$^{-3}$ (Fig. 12). The energy band profiles in both the OFF and ON states show negligible variation between the $10^{16}$ and $10^{17}$ cm$^{-3}$ doping levels (Fig. 12a, b), resulting in comparable transfer characteristics (Fig. 12c). At the

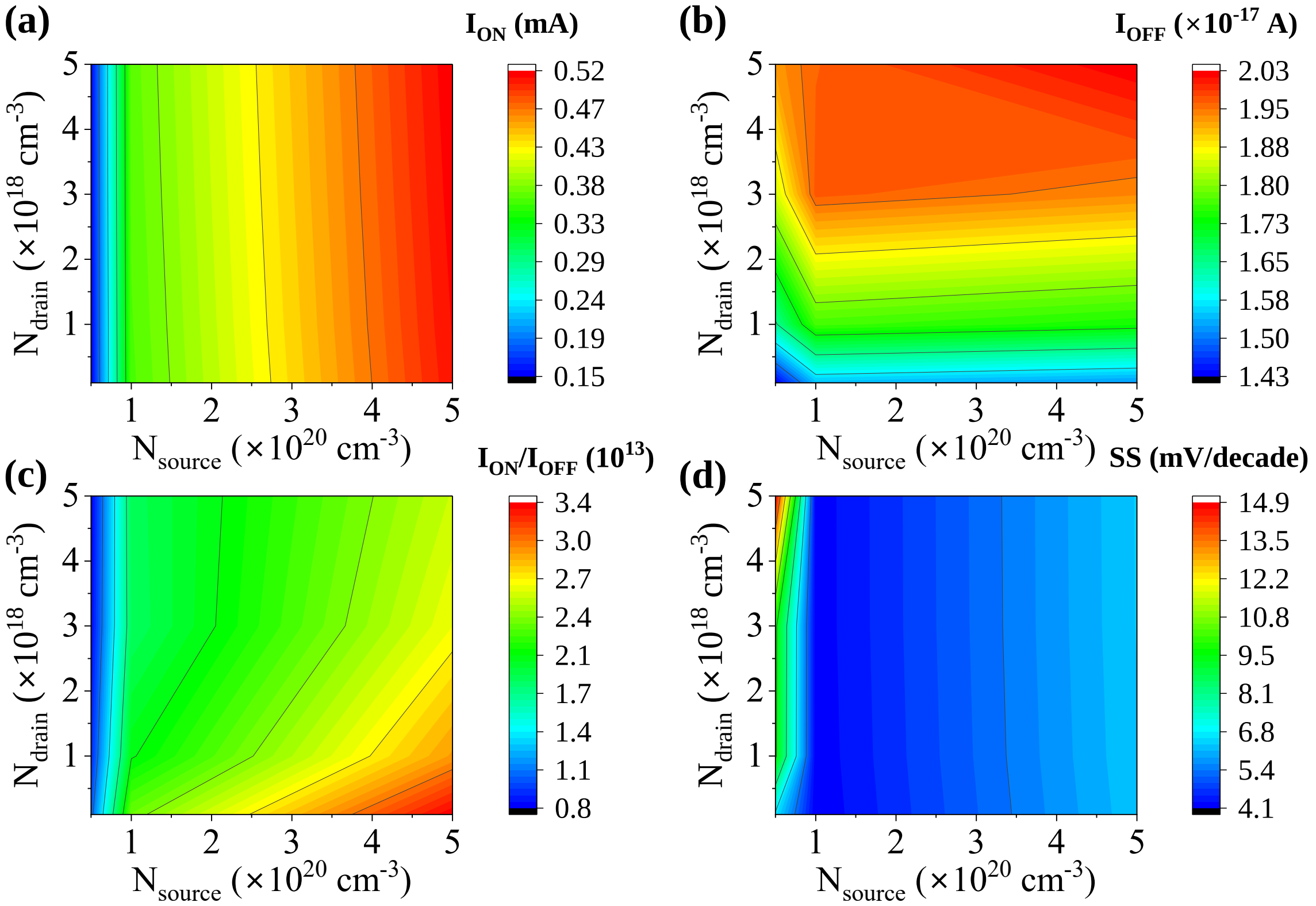

**Fig. 11** (a) $I_{ON}$, (b) $I_{OFF}$, (c) $I_{ON}/I_{OFF}$, and (d) SS for different drain and source doping density.

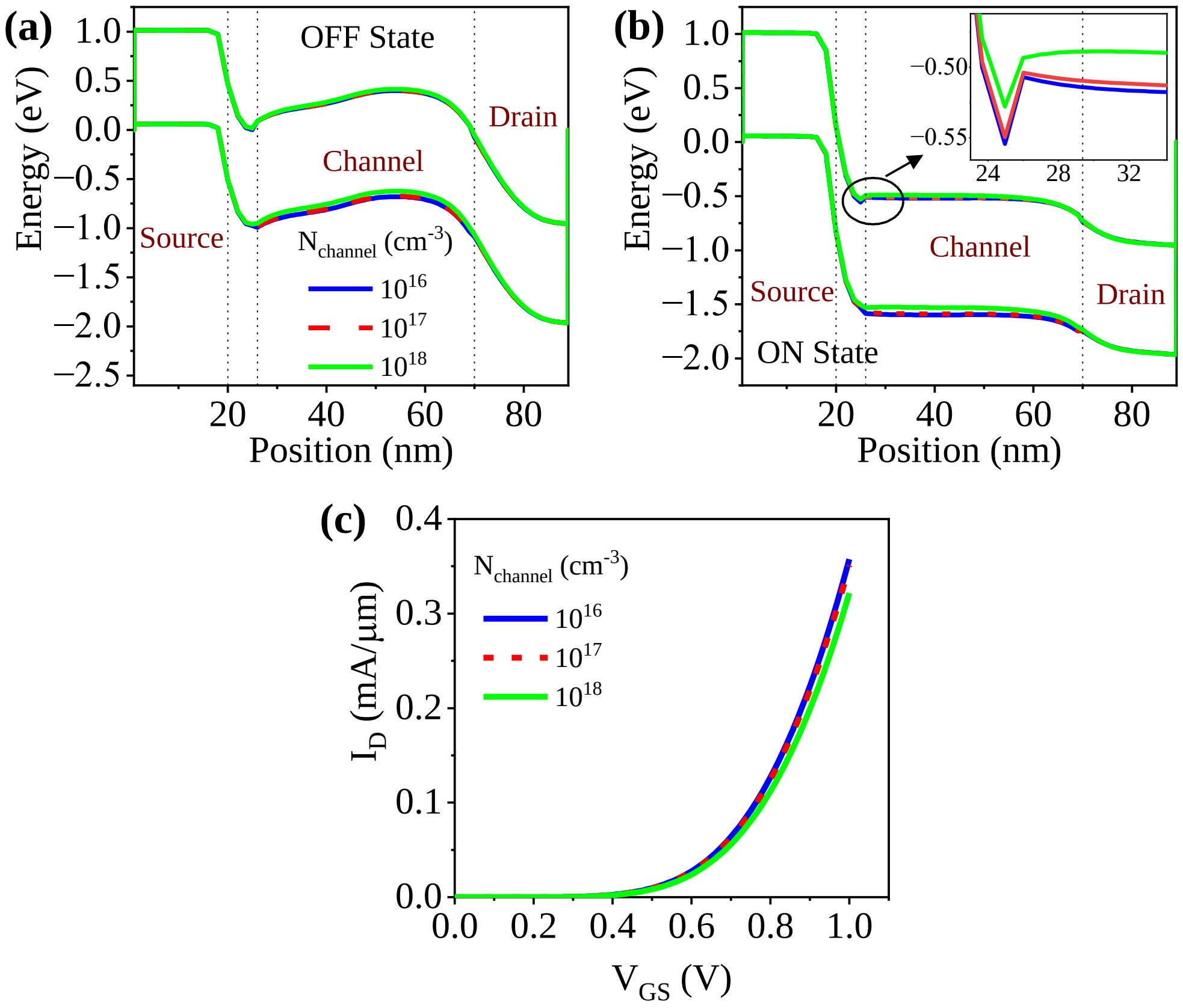

**Fig. 12** Effect of channel doping on: (a) energy band diagram in OFF state ($V_{GS}$ = 0V, $V_{DS}$ = 1V), (b) energy band diagram in ON state ($V_{GS}$ = $V_{DS}$ = 1V), and (c) transfer characteristics. Legends for (b) are the same as for (a).

higher channel doping of $10^{18}$ cm$^{-3}$, the channel becomes more p-type, causing the local $E_C$ minimum within the pocket to shift upward (inset of Fig. 12b). This shift reduces the effective electron tunneling area, leading to a decrease in the ON current (Fig. 12c).

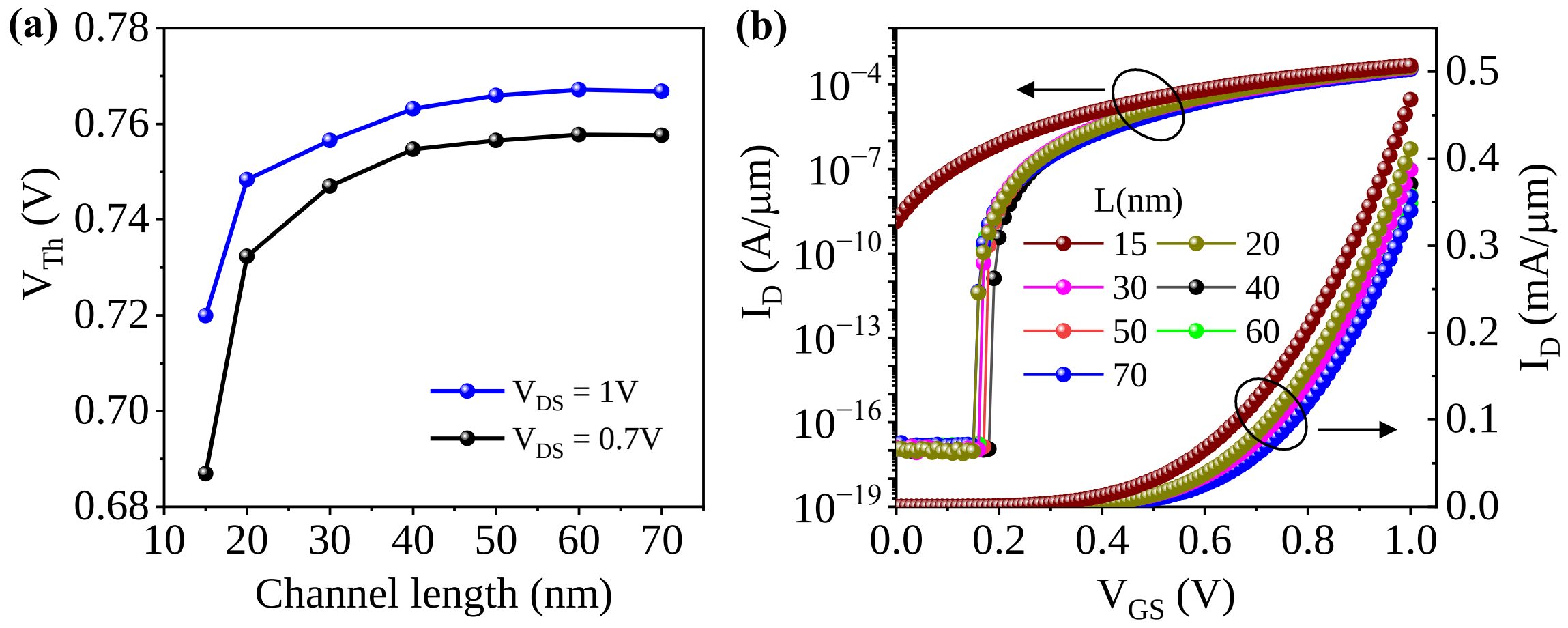

**Fig. 13** (a) Threshold voltage versus channel length and (b) transfer characteristics at different channel length for $V_{DS}$ = 1V.

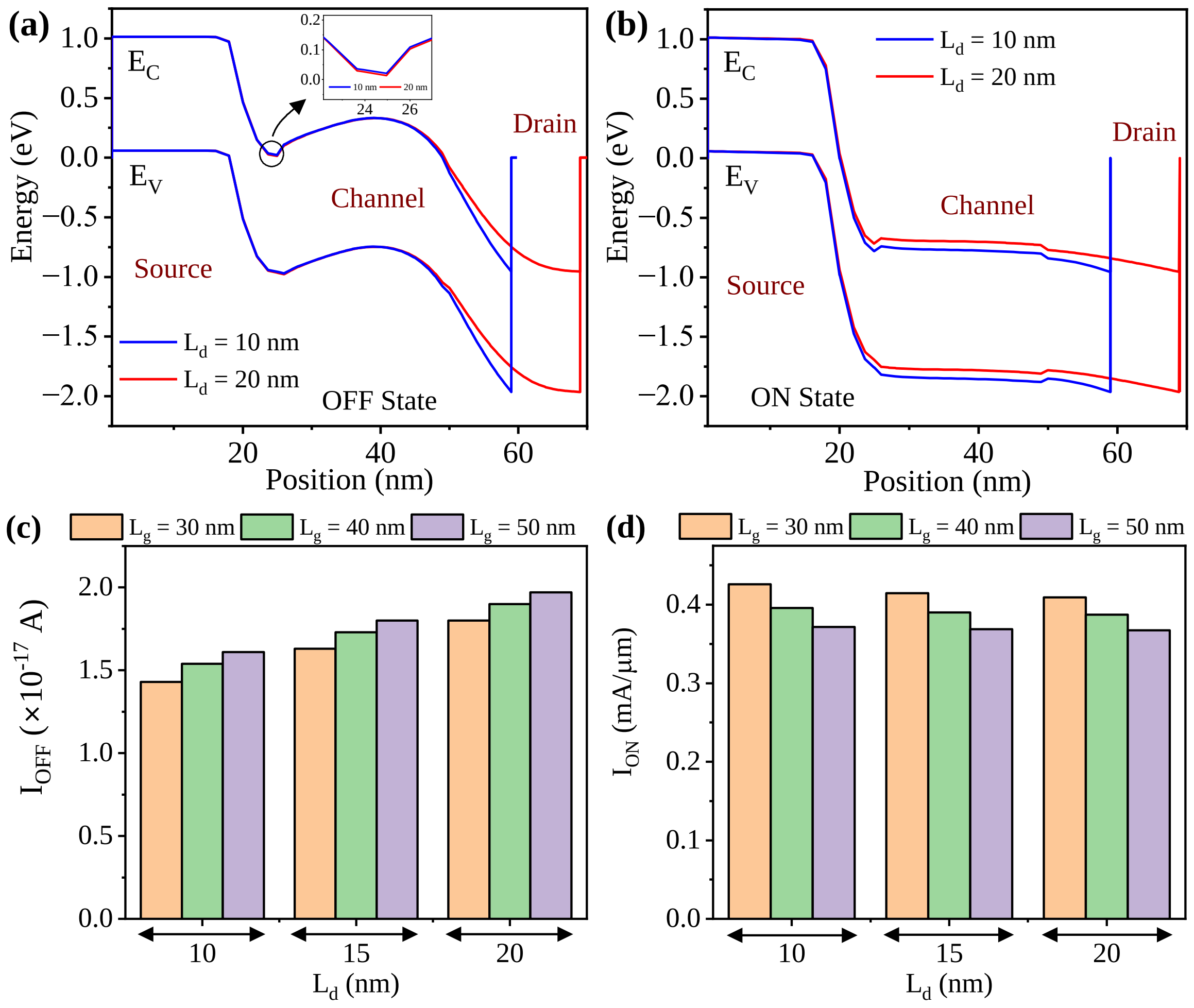

**Fig. 14** Energy band diagram in (a) OFF state ($V_{GS}$ = 0V, $V_{DS}$ = 1V) and (b) ON state ($V_{GS}$ = $V_{DS}$ = 1V) for drain length ($L_d$) of 10 nm and 20 nm at fixed gate length ($L_g$) of 30 nm; (c) OFF current and (d) ON current for different drain and gate lengths.

## 4.4.6 Channel length

The channel length is varied from 70 nm down to 15 nm to investigate short channel effects (Fig. 13). At a fixed drain voltage, the threshold voltage remains nearly constant for channel lengths above 40 nm but exhibits roll-off for lengths below that (Fig. 13a), indicative of enhanced electron BTB tunneling at reduced gate voltages ($V_{GS}$) in short channel devices [39]. Moreover, Fig. 13a demonstrates an increase in threshold voltage with increasing drain voltage ($V_{DS}$), suggesting that the gate maintains quasi-exponential control over the drain current across a wider voltage range at higher $V_{DS}$ [18, 39]. The OFF state current increases significantly when the channel length decreases below 20 nm (Fig. 13b). Simultaneously, the ON current increases with reduced channel length, with the effect becoming more significant for sub-20 nm channel devices (Fig. 13b). In sufficiently short channel devices, the reduced source-to-drain separation leads to significant lateral penetration of the drain bias-induced electric field into the source region, consequently narrowing the tunneling barrier, and enhancing BTB tunneling probability and effective tunneling area [38].

## 4.4.7 Drain length

The drain region length is another important structural parameter that, together with drain doping, determines the distribution of the drain voltage drop and the resulting electric field profile within the channel [38]. This study evaluates the impact of varying drain lengths on the ON and OFF currents for different gate lengths (Fig. 14). In the OFF state, devices with shorter drain lengths exhibit a local $E_C$ minimum positioned at higher energy within the pocket region (Fig. 14a), leading to suppressed OFF current levels (Fig. 14c). Conversely, the ON current increases as the drain length decreases (Fig. 14d), attributable to stronger electric field and enhanced drift velocity of tunneled electrons in shorter drain devices (Fig. 14b). This behavior is more significant in devices with reduced gate lengths.

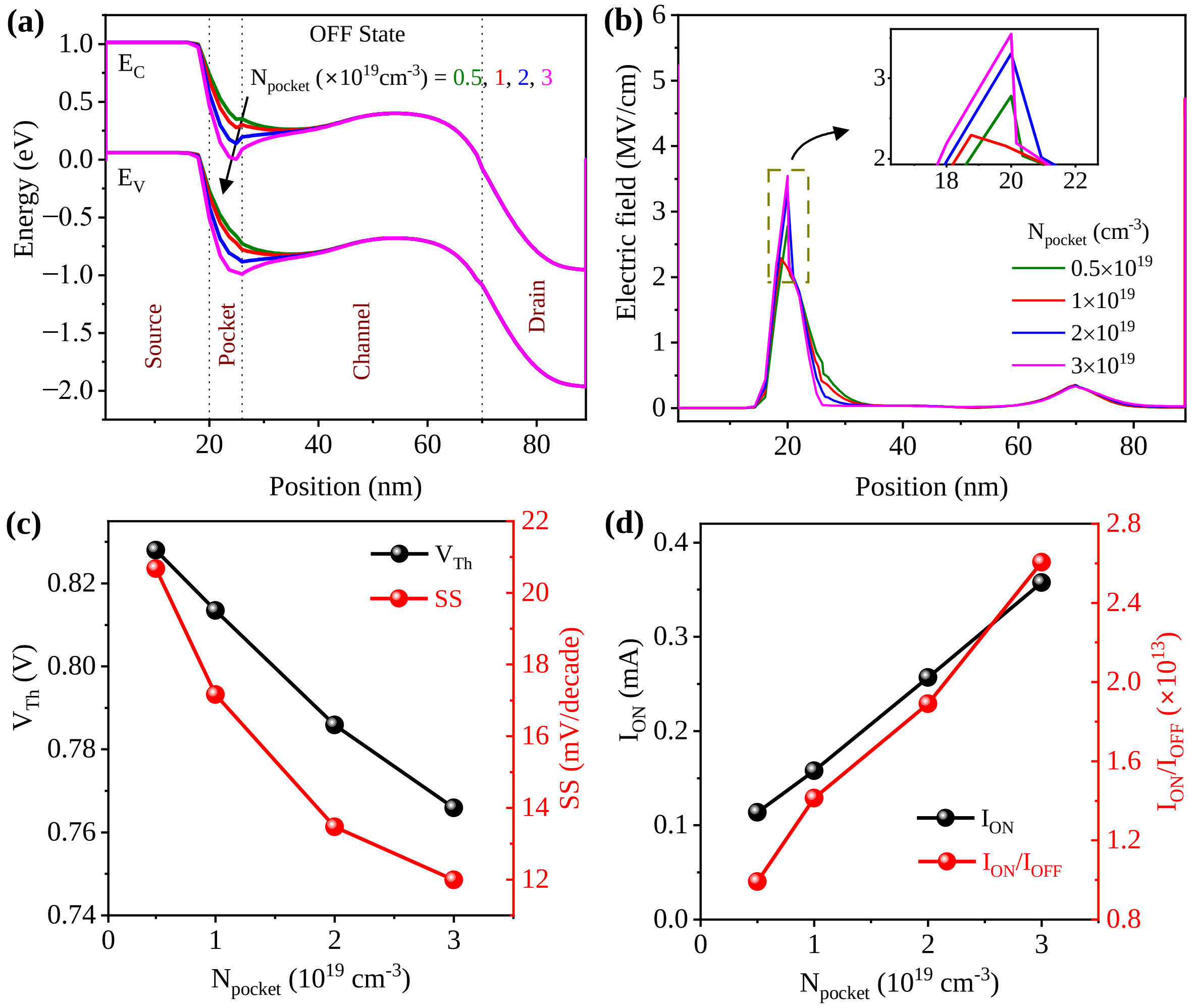

**Fig. 15** Effect of pocket doping density ($N_{pocket}$) on: (a) energy band diagram, and (b) corresponding electric field profile in OFF state ($V_{GS}$ = 0V, $V_{DS}$ = 1V); (c) threshold voltage and subthreshold slope, and (d) ON current and ON/OFF current ratio.

### 4.4.8 Length and doping density of the source-pocket

Figures 15 and 16 illustrate the effects of pocket region doping density ($N_{pocket}$) and length ($L_{pocket}$) on device performance. Initially, $N_{pocket}$ is varied from $5\times10^{18}$ cm$^{-3}$ to $3\times10^{19}$ cm$^{-3}$ while maintaining a fixed $L_{pocket}$ of 6 nm (Fig. 15). An increase in $N_{pocket}$ lowers the local $E_C$ minimum within the pocket (Fig. 15a), which narrows the tunneling barrier and enhances the lateral electric field intensity (Fig. 15b). This facilitates an increased electron BTB tunneling rate [40], resulting in a reduced threshold voltage and increased ON current. Although the OFF current also rises with higher $N_{pocket}$, the relative enhancement of the ON current is more pronounced, yielding an improved ON/OFF current ratio and a reduced subthreshold slope (Fig. 15c, d).

Subsequently, the impact of varying $L_{pocket}$ is assessed at a constant $N_{pocket}$ of $3\times10^{19}$ cm$^{-3}$ (Fig. 16). Both threshold voltage and subthreshold slope decrease with increasing $L_{pocket}$, while ON and OFF currents increase (Fig. 16a, b). The energy band diagrams (Fig. 16c) reveal a downward shift of $E_C$ and a reduction in tunneling barrier width as $L_{pocket}$ increases. This phenomenon is attributed to the possible onset of partial depletion in the pocket region, leading to a higher electron concentration in the undepleted segment (Fig. 16d) [11]. The resultant barrier thinning enhances both ON and OFF currents, with the increase in ON current dominating. Consequently, the ON/OFF current ratio improves, accompanied by a decrease in subthreshold slope.

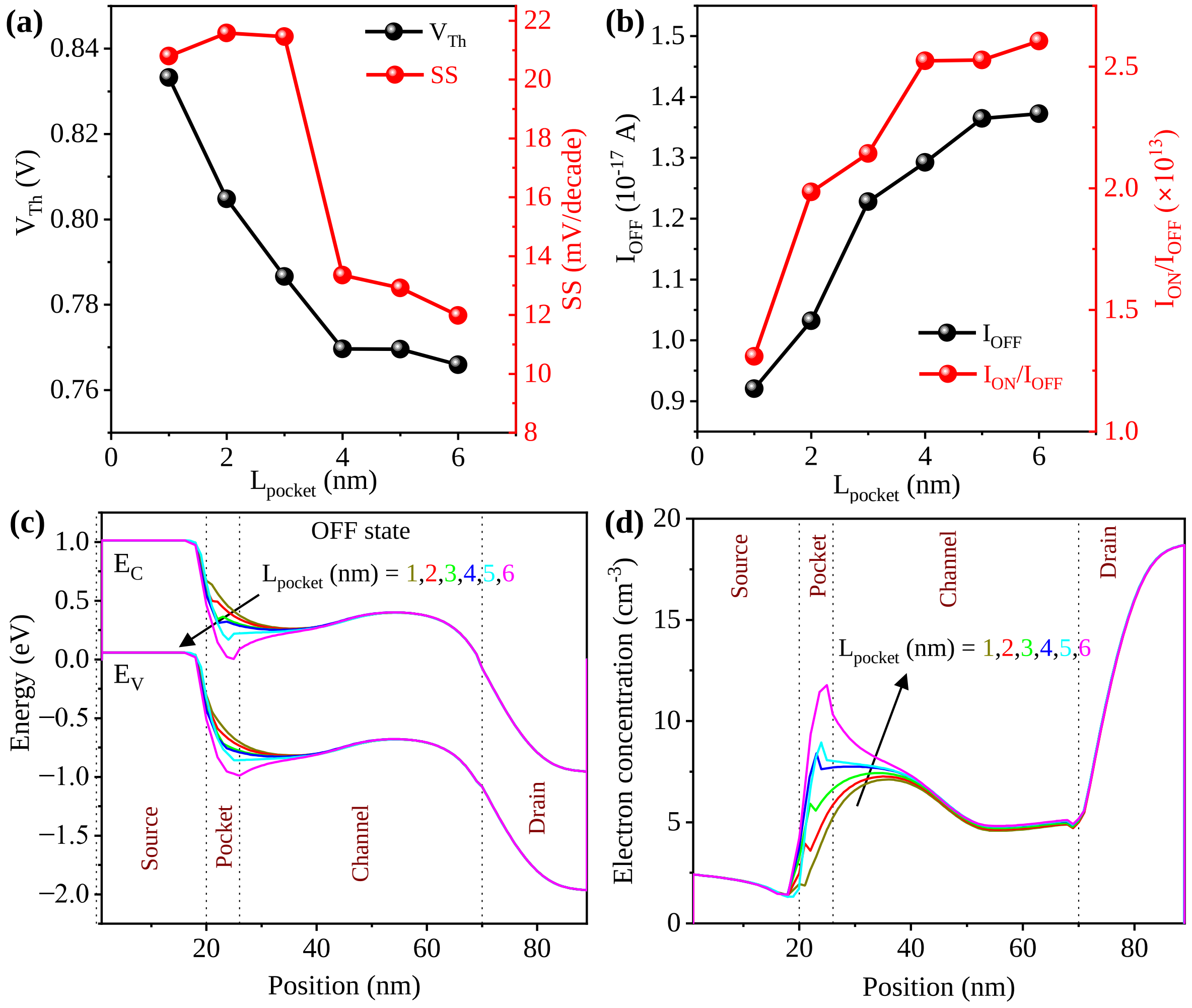

**Fig. 16** Effect of pocket length ($L_{pocket}$) on: (a) threshold voltage and subthreshold slope, (b) OFF current and ON/OFF current ratio, (c) energy band diagram in OFF state ($V_{GS}$ = 0V, $V_{DS}$ = 1V), and (d) corresponding electron concentration profile.

## 4.4.9 Optimized device

Based on the preceding sensitivity analyses of various material and structural parameters, a simulation has been performed at $V_{GS} = V_{DS} = 1V$ using the optimized parameter values: $N_{source} = 5\times10^{20}$ cm$^{-3}$, $N_{drain} = 10^{18}$ cm$^{-3}$, $\Phi_{M1} = 4.3$ eV and $\Phi_{M2} = 4.5$ eV, while all other parameters are retained as specified in Table 1. The simulation results demonstrate excellent performance metrics, including a tunneling barrier width of 3 nm, peak electric field of 3.75 MV/cm at the source-pocket junction, ON current of $3.16\times10^{-4}$ A/μm, OFF current of $1.54\times10^{-17}$ A/μm, ON/OFF current ratio of $2.05\times10^{13}$, and subthreshold slope of 6.29 mV/decade.

# 5 Conclusion

This study utilizes Silvaco Atlas-based 2-D TCAD simulations to investigate the performance of Si-based DMDG-SP TFET featuring a homo-dielectric gate insulator. An initial comparative analysis highlights the advantages of incorporating a source-pocket region and utilizing dual metals for the top and bottom gates. Incorporation of the pocket structure leads to a 6.7x higher ON-state current and a 1.7x lower subthreshold swing relative to the pocket-less device. Additionally, devices with dual-material gates exhibit a 45% increase in ON-state current and a 59% improvement in the ON/OFF current ratio compared to single-material gate counterparts. Subsequently, a comprehensive study with various device parameters reveals key insights: the tunneling gate metal work function primarily modulates threshold voltage, whereas the auxiliary gate metal work function predominantly influences ON current. Increasing the ratio of tunneling to auxiliary gate length enhances tunneling probability and drain current. High-*k* gate dielectrics strengthen gate-to-channel capacitive coupling, resulting in improved ON current and ON/OFF ratio. Higher source doping density reduces the tunneling barrier width and improves tunneling efficiency, while lower drain doping mitigates OFF current, enhances ON/OFF ratio, and suppresses ambipolar conduction. Shorter drain lengths yield higher ON currents and lower OFF currents, and shorter channel length leads to reduced threshold voltage and increased drain current. Higher doping density and optimal length of the pocket region further reduce subthreshold slope and enhance ON/OFF current ratio. Employing these optimized parameters, the device achieves outstanding performance metrics: ON current = $3.16\times10^{-4}$ A/μm, OFF current = $1.54\times10^{-17}$ A/μm, ON/OFF current ratio = $2.05\times10^{13}$, and subthreshold slope = 6.29 mV/decade. These results demonstrate the promising potential of the optimized Si-based DMDG-SP TFET for low-power electronics and next-generation integrated circuit applications.

# Statements and Declarations

**Funding** The authors declare that no funds, grants, or other support were received during the preparation of this manuscript.

**Competing Interests** The authors have no relevant financial or non-financial interests to disclose.

**Author Contributions** All authors contributed equally to the study conception, simulation and analysis, and manuscript preparation.